\documentclass[useAMS,usenatbib]{mn2e}

\usepackage{aas_macros}
\usepackage{epsfig}
\usepackage{amsfonts,amssymb,amsmath}

\usepackage{color}
\usepackage{ulem}
\usepackage{bm}
\usepackage{graphicx}
\usepackage{epsf}

\def\mpch{\,h^{-1}{\rm Mpc}}
\def\msun{\,h^{-1}{\rm M}_\odot}
\def\e{\,{\boldsymbol{e_1}}}
\def\ee{\,{\boldsymbol{e_2}}}
\def\eee{\,{\boldsymbol{e_3}}}

\newcommand{\Rmnum}[1]{\expandafter\@slowromancap\romannumeral #1@}

\title[A general explanation on spin-LSS flip]{A general explanation on the correlation of dark matter halo spin with the large scale environment}
\author[P. Wang \& X. Kang]
{\parbox{\textwidth}{Peng Wang$^{1,2}$\thanks{E-mail:
wangpeng@pmo.ac.cn; kangxi@pmo.ac.cn}, Xi Kang$^{1}$}
\vspace{0.4cm}\\
$^{1}$ Purple Mountain Observatory, the Partner Group of MPI f\"ur Astronomie, 2 West Beijing Road, Nanjing 210008, China\\
$^{2}$ Graduate School, University of the Chinese Academy of Science, 19A, Yuquan Road, Beijing 100049, China}

\begin{document}

\date{Accepted 2017 March 13. Received 2017 March 13; in original form 2016 November 30.}

%\pagerange{\pageref{firstpage}--\pageref{lastpage}} \pubyear{}

\maketitle

\label{firstpage}

\begin{abstract}

Both simulations  and observations have found  that the  spin of halo/galaxy is
correlated  with the large-scale environment, and particularly the  spin of
halo flips in filament. A consistent picture of halo spin  evolution  in
different environments is still lacked. Using N-body  simulation we find that
halo spin with its environment evolves continuously from sheet to cluster, and
the flip of halo spin happens both in filament and nodes.  For the flip in
filament can be explained by halo formation time  and migrating  time when
its environment changes from sheet to filament.   For low-mass haloes,  they
form first in sheets and migrate into filaments later, so their  mass and spin
growth inside filament are lower, and the original spin is still parallel to
filament. For high-haloes, they migrate into filaments first, and most  of
their mass and spin growth are obtained in filaments, so the resulted spin is
perpendicular to filament.  Our results well explain the overall evolution of
cosmic web in the cold dark matter model and can be tested using high-redshift
data.  The scenario can also be tested against alternative models of dark
matter, such as warm/hot dark matter, where the structure formation will
proceed in a different way.

\end{abstract}


\section{Introduction}

It is now  widely accepted that structures in the  universe are formed
from  the  initial seed  of  density  perturbations via  gravitational
instability. On large scale the  structures are characterized as cosmic
web composed of voids, sheets, filaments and nodes (Bond et al. 1996).
The cosmic  web can  be well  described by the  linear theory  and the
Zel'dovich approximation (Zel'dovich 1970).  Dark matter haloes formed
at the  peaks of the density  field, and their properties  are strongly
affected by the large-scale environment. For example, the spin of dark
matter  halo  is mainly  determined  by  the large-scale  tidal  field
(Peebles 1970; White  1984; Porciani 2002).  Halo  shape is also correlated
with the mass distributed on large scales (e.g., van Haarlen \& van de
Weygaert 1993). With the help of N-body simulations, the properties of
dark matter halo and their  relations with large scale structure (LSS)
can be well studied (e.g., Hahn et al. 2007).

The  predicted correlations  of halo  properties with  LSS have  to be
tested  using observations.   Galaxies are  formed within  dark matter
haloes, it is naturally expected  that their properties should also be
correlated with the  LSS. With the advent of large  galaxy sample from
large sky surveys, such  as the SDSS, a lot of  work have been devoted
to  study how  galaxy properties  are correlated  with the  LSS. Among
these studies, the correlation between galaxy spin and the large scale
environment received  particular interests in recent  years (e.g., Lee
\&  Pen 2002;  Trujillo et  al.  2006;  Jones et  al.  2010;  Zhang et
al.  2015).  Most  interestingly,  it is  found  that the  correlation
between galaxy spin and the LSS  is not universal (Tempel \& Libeskind
2013) in the  sense that spin of early type  (or massive) galaxies are
perpendicular to the  nearby filaments, but the spin of  late type (or
low-mass)  galaxies are  parallel to  filaments.  Though  the observed
signal  is very  weak, this  flip of  galaxy spin-LSS  relation is  in
amazing good agreement with  predictions from N-body simulations (Hahn
et  al.  2007;  Aragon-Calvo  et  al.  2007;  Trowland  et  al.  2013;
Libeskind  et  al.   2013;  Forero-Romero  et  al.   2014;  Laigle  et
al. 2015). For a recent review on progress in this field, see Joachimi
et al.  (2015) and the  proceedings of  the IAU Symposium,  Volume 308
(van de Weygaert et al. 2016) .

It is still not clear what is the physical origin for the flip of halo
spin-LSS correlation.   In particular, this  spin flip occurs in
filament environment. In  sheet (sometimes referred as  wall) the spin
is parallel to the sheet  with almost no mass dependence (Aragon-Calvo
et al.   2007; Hahn  et al.   2007). In  nodes (sometimes  referred as
cluster), the  spin of halo  is expected to be perpendicular to the  slowest collapse
direction (see  Section.2 for a  more general description  of spin-LSS
correlation  where the  direction of  LSS is  referred to  the slowest
collapse direction).  Under the assumption that halo spin is converted
from the orbital  angularmentum of accreted mass,  this correlation is
rooted how  halo mass is  accreted.  Indeed, many previous  works have
found that most  subhaloes are accreted along  the filament structures
(e.g., Wang et al. 2005; Libeskind et  al. 2014; Shi et al. 2015; Wang
et al. 2015).  But an universal  direction of mass accretion is unable
to explain the  spin flip of low-mass haloes. Kang  \& Wang (2015) did
find that the  mass accretion respect to the LSS  is not universal but
with a mass  dependence, which can well explain the  flip of halo spin
with the LSS.

It is more intriguing to  understand why the mass flow-LSS correlation
has  a mass  dependence  and how  it  is related  to  the large  scale
environment of  the halo.  Based  on work of  mass flow in  cosmic web
(e.g., Bond  et al.  1996; Sousbie  et al. 2008;  Pichon et  al. 2011;
Cautun et  al. 2014),  a few  studies (Codis et  al.  2012;  Welker et
al. 2014;  Aragon-Calvo et al.  2014;  Pichon et al. 2014)  proposed a
scenario  in which  low-mass  haloes are  formed  by smooth  accretion
through the  winding of flows  embedded in misaligned walls,  so their
spin  are parallel  to the  filaments formed  at the  intersections of
these  walls.   Massive  haloes  are products  of  later  mergers,  in
particular major mergers,  and they acquire spin  perpendicular to the
filament. Such a scenario is  useful to understand the mass dependence
of the spin-LSS  relation.  One major caveat of these  studies is that
they did not  explicitly show the formation and evolution  of the halo
spin-LSS correlation and how they depend on halo mass.

Inspired  by  aforementioned  studies,   here  we   provide  a   more  general
explanation  for  the  evolution  of halo  spin-LSS  correlation  in
different environments, and particularly we show that the flip of halo
spin-LSS  correlation is  mainly determined by  the mass  dependence of  halo
formation time and its migration time  (when the environment of a halo
changes from sheet to filament).   The scenario presented in this work
provides interesting  insight of  hierarchical structure  formation in
the cold dark matter model, and  it can be extended to constrain other
models of  structure formation  or alternatives  of cold  dark matter,
such  as  warm  or  self-interactive  dark  matter,  which  may  predict
different correlations of halo spin with large scale structures.

%%%%%%%%%%%%%%%%%%%%%%%%%%%%%%%%%%%%%%%%%%%%%%%%%%%%%%%%%%%%%%%%%%%%%%%%%
\section{Simulation and Method}
\label{sec:methods}

The simulation data used in this  Letter are the same as those in Kang
\& Wang (2015) in which they studied the accretion of halo mass and its
relation to the large scale environments. We refer the readers to that
paper for details.  Here we give a brief introduction  of the data and
the methods.

The N-body  simulation was run  using the Gadget-2 code  (Spring 2005)
and it follows  $1024^{3}$ dark matter particles in a  periodic box of
$200 \mpch$ with cosmological parameters  from the WMAP7 data (Komatsu
et  al.   2011)  namely:  $\Omega_{\Lambda}=0.73$,  $\Omega_{m}=0.27$,
$h=0.7$ and $\sigma_{8}=0.81$.  The particle masses in this simulation
is  about $5.5\times  10^{8}\msun$ and  60 snapshots  are stored  from
redshift 10  to 0.  At  each snapshot  we identify dark  matter haloes
using the  standard friends-of-friends  (FOF) algorithm (Davis  et al.
1985) with  a link length  that is  0.2 times the  mean inter-particle
separation. To ensure a robust measurement  of halo shape and spin, we
use haloes containing at least 100 particles.  The  full merger  trees
of each halo are also constructed (see Kang et al. 2005 for details).

The spin  $\boldsymbol{J}$ of each FOF halo is obtained as,
%=============================================================
\begin{equation}
  \boldsymbol{J}=\sum_{i=1}^N m_i \boldsymbol{r_i} \times (\boldsymbol{v_i}-\boldsymbol{\bar{v}}),
\label{equ:am}
\end{equation}
%=============================================================
respectively, where $m_i$ is  particle mass, $\boldsymbol{r_i}$ is the
position  vector of  particle  $\boldsymbol{i}$ relative  to the  halo
center,  $\boldsymbol{v_i}$  is  the  velocity of  the  each  particle
$\boldsymbol{i}$ and  $\boldsymbol{\bar{v}}$ the mean velocity  of the
halo.  The LSS  environment  around  a halo  is  determined using  the
Hessian  matrix of  the smoothed  density field  at the  halo position
(e.g.,  Hahn  et  al.  2007;  Arag'on-Calvo  et  al.  2007;  Zhang  et
al. 2009), which is defined as:
%=============================================================
\begin{equation}
  H_{ij}=\frac {\partial^2\rho_s(\boldsymbol{x})} 
  {\partial x_{j} \partial x_{j}},
\label{equ:hessian}
\end{equation}
%=============================================================
where  $\rho_s(\boldsymbol{x})$ is  the  smoothed  density field  with
smoothing  scale $R_s$.   The eigenvalues  of the  Hessian matrix  are
sorted as $\lambda_{1}<\lambda_{2}<\lambda_{3}$, and the corresponding
eigenvectors    are   labeled    as    $\e$,    $\ee$   and    $\eee$
respectively. Depending on the number of positive eigenvalues, the LSS
environment  is  classified as  node,  filament,  sheet or  void.   As
explained in Kang  \& Wang (2015), $\eee$ is the  direction of slowest
collapse of mass  on large scale. As found by  Libeskind et al. (2014)
and Kang \& Wang (2015), $\eee$  is a good and universal definition of
the direction of LSS.  For instance, in a filament environment, $\eee$
is  the filament  direction, and  for a  sheet environment,  $\eee$ lies in the plane of the sheet. In node region,
$\eee$   indicates  the   latest  collapsed   and  least   compressed
direction.    As     in    Kang    \&    Wang     (2015),    we    use
$\cos\theta_3=\boldsymbol{j}  \cdot \eee$  to  indicate the  alignment
between halo spin and the LSS. For a random distribution, the expected
$<\cos(\theta_3)>$ is  0.5, and  if $<\cos(\theta_3)>$ is  larger than
0.5,  the halo  spin is parallel  to  the LSS,  and  otherwise it is
perpendicular to the LSS.

For each halo with mass $M_0$ at $z=0$, we trace its formation history
using  its  main progenitors.  Following  the  common used  definition
of halo formation time (e.g., Sheth \&  Tormen 2004), $z_f$, is the
time when mass of the main progenitor, $M(z_f)$, is around half of
the mass at  $z=0$, $M(z_f) \sim 0.5M_0$. From  the evolution scenario
of cosmic web (Zel'dovich 1970; Bond  et al. 1996), the environment of
a halo usually follows the sequence  from sheet, filament and to
node, and during this progress the  halo grows accordingly through mass
accretion in the cosmic web.   For  a halo in a filament environment at
$z=0$, we trace the evolution of its environment,  and we define
an entering time, $z_e$, when the environment of a halo finally changes
(most of filament haloes have only one $z_e$, but the outliers
may have more then one $z_e$, for outliers we select the smallest (cloest to z = 0)
one as their entering redshift). 
from  sheet to  filament. We  find that  for most  haloes in  filament
environment  at $z=0$,  their environment  did not  change after  they
enter into  filament since $z_f$. For
the results below,  we will show the  alignment signal, $\cos\theta_3$,
for haloes at  their formation, entering times and compare  them to the
alignment at $z=0$ to see how the evolution depends on halo mass.

%%%%%%%%%%%%%%%%%%%%%%%%%%%%%%%%%%%%%%%%%%%%%%%%%%%%%%%%%%%%%%%%%%%%%%%%%
\section{Results}
\label{sec:results}

\begin{figure*}
  \includegraphics[width=0.85\textwidth,height=0.3\textheight]{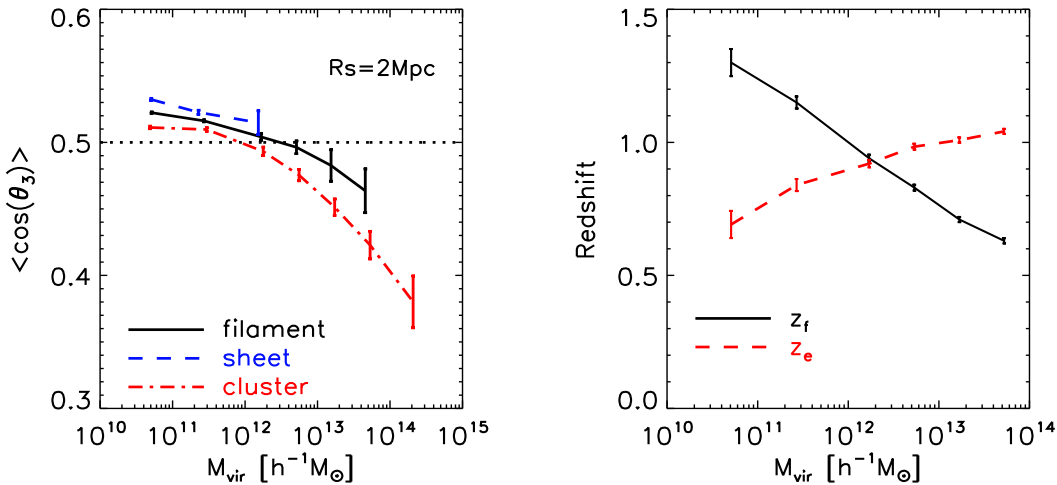}
  \caption{Left panel: The  mean $\cos(\theta_3)$ of halo  spin-LSS as a
  function  of halo  mass at  redshift $z=0$.  Here $\theta_3$  is the
  angle  between halo  spin vector  and the  direction of  large scale
  structure.  Different color  lines with
  error bars are for haloes  in different environments at $z=0$. Right
  panel:  The  formation and  entering  time  for haloes  residing  in
  filaments  at $z=0$.  The entering  time  is the  redshift when  the
  environment  of the  main progenitor  of a  $z=0$ halo  changes from
  sheet to filament, see text for more detail.  }
\label{fig:fig1}
\end{figure*}

\subsection{Halo spin-LSS correlation in different environment}

As a  first step we show  the halo spin -LSS  correlation in different
environment at  $z=0$ in the  left panel of  Fig.~\ref{fig:fig1}.  The
blue dashed line for haloes  in sheets, black solid for filaments, and
red  dash-dot line for  cluster.  In  agreement with  previous results
(e.g., Hahn et al. 2007; Aragon-Calvo et al. 2007; Zhang et al. 2009),
the halo spin in sheets are parallel to $\eee$, and in filaments there
is a mass dependence.  The flip of spin-LSS occurs at haloes with mass
around  $3 \times  10^{12}\msun$, also  agrees with  other results
(e.g., Codis et  al. 2012). 
For the first time we show that in nodes, the spin of halo with
mass $<5\times10^{11}\msun$ is align with $\eee$, but turning into
perpendicluar to $\eee$ for more massive haloes. For haloes with mass bigger
than flip mass, the alignment signal with more stronger mass dependence than in
other environments.

The  evolution  of  halo  spin-LSS  correlation  from  sheet  to  cluster
environment can be understood as a consequence of cosmic web formation (e.g.,
Bond et  al. 1996; Codis  et al. 2012; Pichon  et al. 2015):
in sheet environment, the mass collaspe along the  $\e$
direction is fastest (secondly along $\ee$, slowest along $\eee$), so mass flow
is mostly from $\e$ and partly form $\ee$ to  the sheet plane ($\ee$-$\eee$
plane), and the resulted spin is  in the  sheet plane  and being  parallel to
$\eee$.  With  further collapse along $\ee$, a filament begins  to form, and
the mass flow is mainly along $\ee$  and the halo spin is still  parallel to
$\eee$. At the final  stage, when  the mass  along $\eee$  is collapsed  with
formation of a node environment, the  mass flow is mainly along $\eee$ and the
resulted spin is then being perpendicular to $\eee$.

The above scenario  is useful to understand the  overall evolution of the
spin-LSS  correlation in different  environments, but it  is still not clear
why in  filament and cluster environment the  spin of low-mass
halo is parallel to filament, but  being perpendicular to $\eee$ for massive
halo. It is obvious that in  filament, the anisotropy of mass flow has a
halo-mass dependence.  

We mainly focus on the spin flip in filament in this $Letter$.
It is naturally expected that  this could be related to how long a halo  has
stayed in the filament. To investigate this, in the right panel  of
Fig.~\ref{fig:fig1} we show the formation time and entering  time for haloes in
filaments at  $z=0$.  It is seen that the formation  time of halo (black solid
line) is mass dependent that low-mass halo forms first,  in agreement with the
CDM predictions (e.g.,  Navarro et  al. 1997).   But the  red dashed  line
shows that low-mass haloes  enter filaments  later than larger  haloes.
Basically low-mass haloes enter filaments after they are formed ($z_f>z_e$),
but high-mass haloes  enter filaments before they  are formed ($z_e>z_f$).  The
intersection  between formation time  and entering time  occurs at halo  with
mass  around  $3\times 10^{12}\msun$,  very  close  the transit mass of
spin-LSS relation in filament, and  also close to the characteristic mass of
collapsing halo at $z=0$.

\subsection{The evolution of halo spin-LSS correlation}

The right  panel of  Fig.~\ref{fig:fig1} has important  implication on the
origin of the flip of halo spin-LSS correlation, it indicates that halo
spin-LSS correlation  could be  linked to  the time  a halo  has stayed in
filament. To shed light on it, we divide haloes in filaments at  $z=0$  into
two  sub-samples  based  on their  virial mass, one  is those
with mass smaller than flip mass, $M_{flip}$, labeled  as $align$ sample, the
other is those with mass bigger than $M_{flip}$, labeled as $perp$ sample.
Basically the $align$  sample are dominated by low-mass haloes and $perp$
sample are dominated by high-mass haloes.

\begin{figure*}
  \includegraphics[width=0.85\textwidth,height=0.3\textheight]{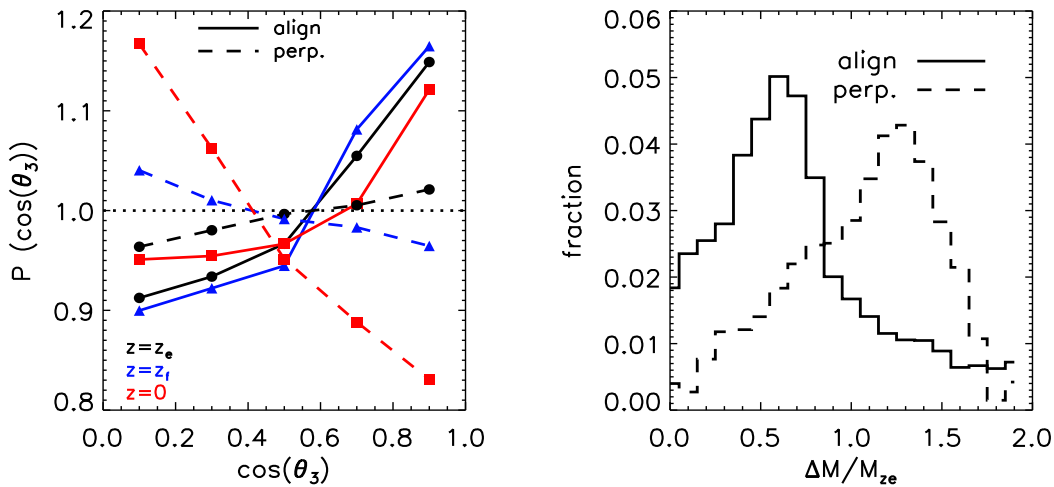}
  \caption{
Left panel: The probability distribution of $\cos\theta_3$ at different
redshifts for two sub-samples.  Here we plot the spin-LSS correlation at three
redshifts: $0$, $z_e$ (entering time) and $z_f$ (formation time). The value
$P=1$ means a random distribution between halo spin vector and the direction of
large scale structure, $\eee$.  Right panel: the distribution of mass growth
rates for the two sub-samples, here  $\Delta M=M_0-M_{z_e}$, in which
$M_{z_e}$, $M_0$ is the virial mass of the progenitor halo at redshift $z=z_e$,
and the mass of the descendant halo at $z=0$.
        }
\label{fig:fig2}
\end{figure*}

In the left panel of Fig.~\ref{fig:fig2} we show the time evolution of
the  spin-LSS correlation  for  the two  sub-samples  using solid  and
dashed  lines.  For  each  sub-sample, we  plot  the  distribution  of
$\cos(\theta_3)$ of their  progenitors at the time  of formation (blue
lines), time  of entering  into filament  (black lines).   Remind that
both the formation time $z_f$ and  entering time $z_e$ of each halo is
determined from its  formation history.  For comparison,  we also plot
the distribution of $\cos(\theta_3)$ for  the two sub-samples at $z=0$
with red lines.

Fig.~\ref{fig:fig2}  shows an interesting  evolution pattern.  For the
$align$  sample, the  blue  solid line  shows  that the  times of  their
formation, their spin is parallel to  the LSS, and when they enter the
filament, their  spin is still  parallel to the filament,  but becomes
slightly  weaker  (black solid  line).  At  $z=0$ their  spin-filament
alignment  continuously  become  weaker,  but  still  being  align  with
filaments (red solid  line). However, for the $perp$  sample, they enter
filament  firstly, with spin  parallel to  the filament  (black dashed
line).   With  mass  grow   in  the   filament,  their   spin  becomes
perpendicular to filament at time  of formation (blue dashed line). At
$z=0$  the  mis-alignment  signal  becomes more  stronger,  and  being
perpendicular to the filaments (red dashed line).

In the right panel of Fig.~\ref{fig:fig2}  we show the mass growth for
the two  sub-samples of haloes.  It  is clearly seen that  for $align$
haloes (solid  line), their  mass growth  after entering  filaments is
lower and  most haloes  increase their  mass by  less than  80\%.  For
$perp$  haloes (dashed  line), their  mass growth  inside filament  is
significant, and  most of them  doubles their mass after  entering the
filament.   Fig.~\ref{fig:fig2} clear  shows  that for  all haloes  in
filament at $z=0$,  their initial spin is parallel to  filament at the
time when they enter filament, but  the fate is different for low-mass
and  high-mass haloes.   For low  mass haloes,  most of  them mass  is
assembled before they enter into  filament, although they accrete mass
in  filament  environment,  their  spin   is  still  parallel  to  the
filament. For  massive haloes,  their mass  growth inside  filament is
significant, so the  mass accretion along filament led  to their final
spin to be perpendicular to filament.

\section{Conclusion and Discussion}
\label{sec:con}

In this  Letter, we investigate the  origin of the mass  dependence of
halo spin  with its environment.  We find  that there is  an evolution
pattern  in the  halo  spin-LSS correlation  in different  large-scale
environment that in sheets halo spin is parallel to LSS with almost no
mass dependence.
The halo  spin-LSS correlation  in filaments and nodes
has a  mass dependence that low-mass  haloes have
their  spin  being parallel  to  $\eee$,  and high-mass  halo  being
perpendicular  to  $\eee$.  Such   an  evolution  of  halo  spin-LSS
correlation  in different  environment  agrees  with previous  studies
(e.g., Hahn et al. 2007; Codis et al. 2012).

We focus on the physical origin of halo spin in filament.
We find that the mass dependence of halo spin-LSS  is due to the mass
dependence of halo formation  time and halo migrating time from sheet  to
filament.  For  low-mass  haloes, they  forms first  with spin  being
initially parallel to filament. After they enter into filament,their mass
growth is minor, hardly changing their spin direction.  For high-mass haloes,
they firstly enter  filament also with spin parallel  to filament, but due to a
significant mass growth  inside filament, their final spin is perpendicular to
filament.   The scenario presented in  this letter is in agreement with the
scenario  proposed by other studies (e.g., Codis et  al.  2012;  Welker  et
al.   2014;  Pichon  et  al.   2015).  Our explanation is more general by
connecting halo formation and migrating time which can be further tested using
observational data. 
In addition, we also find the the conclusion still holds for haloes in cluster region. 
For more detals, we will discuss in our next paper.

Although observational test of simulated  prediction of halo spin with
the large scale  environment is not easy, recent  progress using large
sky survey  make it possible, such  as those studies by  a few authors
(e.g., Zhang et  al. 2014). In particular, Tempel  \& Libeskind (2013)
confirmed  the  mass  dependence  of  galaxy  spin  with  large  scale
structure  using local  observations.  Our  results suggests  the halo
spin-LSS  is  evolving, depending  on  halo  formation time  and  mass
accretion  history, it  is thus  expected that  such a  correlation is
dependent on galaxy properties, such as color, star formation rate,
and the correlation will also  evolve with redshift.  With more survey
data  available at  high  redshifts in  future,  the predicted  galaxy
spin-LSS can  be further tested,  as achieved  by a recent  work using
VIMOS survey (Malavasi et al. 2016).

On  the  other hand,  our  results  in this  paper  are  based on  CDM
simulation,  and  it  is  well   know  that  dark  matter  halo  forms
hierarchically in  CDM and  the environment changes  subsequently from
sheet to filament  and to node finally.  Current  results and proposed
scenario support the  CDM model.  However, if the dark  matter mass is
not  cold, such  as hot  or warm  dark matter,  the formation  of dark
matter  halo  and  its  relation  to environment  will  proceed  in  a
different way (e.g.,  Colombi et al. 1996; Knebe et  al. 2003; Bode et
al. 2001;  Frenk \& White  2012).  In that  case our scenario  of halo
spin-LSS  correlation   will  predict  a  different   correlation  and
evolution.   More studies  can  be done  using  N-body simulations  to
investigate the predictions in other models of dark matter particle or
structure formation scenario.

\section{Acknowledgments}
The  work   is  supported  by  the  973   program  (No.  2015CB857003,
No. 2013CB834900), NSF of  Jiangsu Province (No. BK20140050), the NSFC
(No.  11333008)  and  the  Strategic  Priority  Research  Program  the
emergence of cosmological structures of the CAS (No. XDB09010403). The
simulations are run on the supercomputing center of PMO.

\end{document}